\documentclass[a4paper,11pt]{article}

\usepackage{jheppub}
\usepackage{multirow}
\usepackage{xspace}

\newcommand{\Rdphi}{R_{\Delta \phi}}

\newcommand{\Dphimax}{\Delta \phi_{\rm max}}
\newcommand{\Dphi}{\Delta \phi_{\rm dijet}}

\newcommand{\mur}{\mu_R}
\newcommand{\muf}{\mu_F}

\newcommand{\as}{\alpha_s}
\newcommand{\ass}{\alpha_s^2}
\newcommand{\asss}{\alpha_s^3}
\newcommand{\assss}{\alpha_s^4}
\newcommand{\asmz}{\alpha_s(M_Z)}

\newcommand{\Rcone}{R_{\rm cone}}

\newcommand{\ord}{{\cal O}}
\newcommand{\ppbar}{p{\bar{p}}}

\newcommand{\ptmax}{\ensuremath{p_{\rm T max}}\xspace}
\newcommand{\ptmin}{\ensuremath{p_{\rm T min}}\xspace}
\newcommand{\yinit}{\ensuremath{y_{\rm initial}}\xspace}
\newcommand{\yfinal}{\ensuremath{y_{\rm final}}\xspace}
\newcommand{\nlojet}{{\sc NLOJet++}\xspace}
\newcommand{\fastnlo}{{\sc fastNLO}\xspace}

\newcommand{\ie}{\mbox{i.e.}\xspace}     
\newcommand{\eg}{\mbox{e.g.}\xspace}     
\newcommand{\pt}{\ensuremath{p_\mathrm{T}}\xspace}
\newcommand{\DO}{\mbox{D\O}\xspace}

\title{\boldmath Dijet azimuthal decorrelations for $\Dphi < 2\pi/3$
                 in perturbative QCD}

\author{M. Wobisch,}
\author{K. Rabbertz}

\affiliation[a]{Department of Physics, Louisiana Tech University,\\
  600 Dan Reneau Dr., Ruston, LA, USA}
\affiliation[b]{Institut f\"ur Experimentelle Kernphysik, KIT,\\
  Postfach 6980, D-76128 Karlsruhe, Germany}

\emailAdd{wobisch@latech.edu}
\emailAdd{klaus.rabbertz@cern.ch}

\abstract{We point out an inconsistency in perturbative QCD
  predictions previously used for dijet azimuthal decorrelations for
  azimuthal angles of $\Delta\phi_{\rm dijet} < 2\pi/3$ between the
  two jets. We show how the inconsistency arises and how the
  calculations can be modified to provide more accurate results that
  exhibit a smaller scale dependence and give a better description of
  the data than the inconsistent results. We also
  explain how the quality of the predictions strongly depends on a
  perceivedly minor detail in the definition of the dijet phase space
  and give recommendations for future measurements.
}


\begin{document}
\maketitle
\flushbottom

\section{Introduction\label{sec:intro}}

Measurements of dijet azimuthal decorrelations in hadron-hadron
collisions provide a unique testing ground for the predictions of
perturbative quantum chromodynamics (pQCD).  The dijet azimuthal
decorrelation studies the production rates of dijet events as a
function of the azimuthal angular separation between the two
jets in an event that define the dijet system, $\Dphi = | \phi_{\rm
  jet1} - \phi_{\rm jet2} |$.  The measured quantity, labeled $P$ in
this article and originally proposed by the \DO
collaboration~\cite{Abazov:2004hm}, is the dijet differential cross
section, ${\rm d} \sigma_{\rm dijet}/{\rm d} \Dphi$, normalized by
the inclusive dijet cross section, $\sigma_{\rm dijet}$, integrated
over $\Dphi$:

\begin{equation}
  P \, = \, \frac{1}{\sigma_{\rm dijet}} \cdot
  \frac{{\rm d} \sigma_{\rm dijet}}{{\rm d} \Dphi}
  \,.
  \label{eq:def}
\end{equation}

The range of kinematically accessible values in $\Dphi$ is indicated
in figure~\ref{fig:sketch} for processes with final states of
different jet multiplicities.  In $2 \rightarrow 2$ processes $\Dphi$
has always the largest possible value of $\Dphi = \pi$
(figure~\ref{fig:sketch}~a).  If $\Dphi$ is significantly below $\pi$,
then the quantity $P$ is probing hard $2 \rightarrow 3$ and 
$2 \rightarrow 4$ processes, \ie three-jet and four-jet production.
Following the \DO measurement, the quantity $P$ was also measured by
the CMS and ATLAS
collaborations~\cite{Khachatryan:2011zj,daCosta:2011ni}. In all
measurements the data are fairly well described by the theory
predictions at next-to-leading order (NLO) pQCD for 
$3\pi/4 \lesssim \Dphi < \pi$.
For smaller $\Dphi$, in particular for $\Dphi <2\pi/3$, the
theory predictions exhibit a large renormalization scale dependence and
lie significantly below the data.

In this article, we focus on the comparison of fixed-order pQCD
predictions and data in the kinematic region of $\Dphi < 2\pi/3$.
In section~\ref{sec:ps} we introduce and compare the phase space definitions in
the different analyses and discuss their effects on the kinematic
constraints in $2 \rightarrow 3$ processes.
In section~\ref{sec:pqcd} we show that the pQCD calculations by two of the
experimental collaborations~\cite{Abazov:2004hm,Khachatryan:2011zj}
for the region of $\Dphi < 2\pi/3$ are inconsistent, and demonstrate
how a correct treatment provides pQCD
predictions with a reduced scale dependence. The results of these
calculations also give a better description of the experimental data, as
shown in section~\ref{sec:results}.
In section~\ref{sec:future} we discuss how a particular choice in the
selection of the dijet phase space in the third experimental
analysis~\cite{daCosta:2011ni} renders fixed-order pQCD predictions less
accurate and how this can be improved in future measurements by
a small modification in the dijet phase space definition.

\section{Phase space and kinematic constraints\label{sec:ps}}

\begin{table}[tbp]
\centering
\begin{tabular}{|c|c|c|c|}
\hline
 & \multicolumn{3}{c|}{Experiment (reaction and center-of-mass energy)} \\
Parameter
 & \DO  ($\ppbar$, 1.96\,TeV ) & CMS ($pp$, 7\,TeV) & ATLAS ($pp$, 7\,TeV)\\
\hline
jet algorithm &  Run II cone  & anti-k$_t$    & anti-k$_t$ \\
jet radius & $\Rcone=0.7$ & $R=0.5$    & $R=0.6$ \\
$\yinit$      & $\infty$  & 5.0 & 2.8 \\
$\yfinal$  &  0.5      &      1.1 &      0.8 \\
$\ptmin$  & 40\,GeV   & 30\,GeV  & 100\,GeV \\
$\ptmax$ ranges &   75--100\,GeV &  80--110\,GeV  &   110--160\,GeV \\
                &  100--130\,GeV & 110--140\,GeV  &   160--210\,GeV \\
                &  130--180\,GeV & 140--200\,GeV  &   210--260\,GeV \\
                &    $>$180\,GeV & 200--300\,GeV  &   260--310\,GeV \\
                &                &   $>$300\,GeV  &   310--400\,GeV \\
                &                &                &   400--500\,GeV \\
                &                &                &   500--600\,GeV \\
                &                &                &   600--800\,GeV \\
                &                &                &   $>$800\,GeV \\
\hline
\end{tabular}
\caption{\label{tab:ps}Summary of the parameters defining the dijet
  phase space in the \DO, CMS, and ATLAS measurements
  of dijet azimuthal decorrelations~\cite{Abazov:2004hm,
    Khachatryan:2011zj, daCosta:2011ni}.
  Variables are defined in the text.}
\end{table}

\begin{figure}
  \centering
  \includegraphics[scale=1.1]{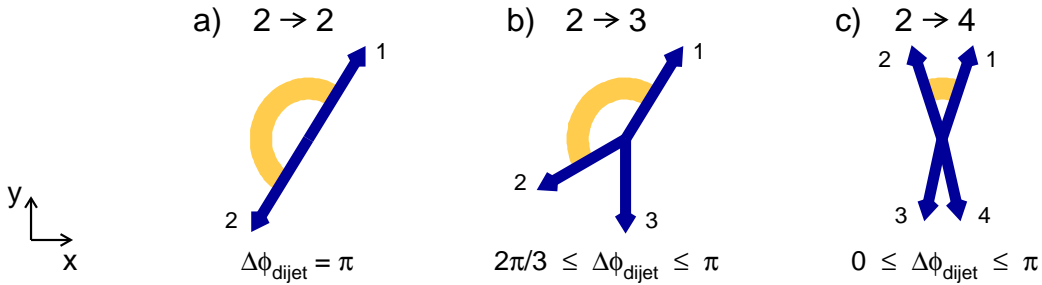}
  \caption{\label{fig:sketch}Sketches of the azimuthal angular separation
    $\Dphi$ between the two jets leading in $\pt$ in an event for
    $2\rightarrow 2$, $2\rightarrow 3$, and $2\rightarrow 4$ processes.
    Also indicated is the kinematically accessible range in $\Dphi$
    for the three configurations.}
\end{figure}

For a given process (\eg $pp$ or $\ppbar$ collisions) and
center-of-mass energy, the measured quantity $P$, defined
in equation~(\ref{eq:def}),
depends on additional choices, including the jet algorithm with its
parameters, and the requirements on the jet rapidities $y$ and the
transverse jet momenta $\pt$ with respect to the beam direction.  The
initial jet selection may be carried out in a limited $y$ region, with
$|y| < \yinit$ (where $\yinit$ can be adapted to the detector
acceptance).  The dijet system is then defined by the two jets with
the highest $\pt$ inside this region; here, these are labeled ``jet1''
and ``jet2''.  The final phase space for the rapidities $y_{1,2}$ of
jet1 and jet2 is then further constrained by $|y_{1,2}| < \yfinal$.
Furthermore, the $\pt$ of jet2 is required to be above a given threshold,
$\ptmin$, and the analysis results are presented in different regions
of the $\pt$ of jet1, $\ptmax$.
An overview of the choices for these parameters in the analyses by the
\DO, CMS, and ATLAS experiments is given in table~\ref{tab:ps}.
The main difference between the three scenarios regarding the
scope of this article is the choice of $\yinit$. In
the \DO scenario, the $y$ region for the initial jet selection is
unlimited ($\yinit = \infty$), while the ATLAS and CMS scenarios are
limited to $\yinit = 2.8$ and $5.0$, respectively~\cite{daCosta:2011ni,Rabbertz:2011pc}.
As a consequence of the choices for $\yinit$ and $\ptmin$,
the three scenarios then have different kinematic constraints for $2
\rightarrow 3$ processes as explained below:

\begin{itemize}

\item {\bf Kinematic constraints for an unlimited \boldmath $y$
    region,    $\yinit = \infty$}\\
  For $\yinit = \infty$, the selected jets, jet1 and jet2, are always
  the two jets leading in $\pt$ of the entire event. This selection
  criterion results in the kinematic constraint that the smallest
  possible $\Dphi$ value in a $2 \rightarrow 3$ process (\ie in a
  three-jet final state) is $\Dphi = 2\pi/3$ (cf.\
  figure~\ref{fig:sketch}~b), while angles of $\Dphi < 2\pi/3$ are
  only accessible in final states with four or more jets (cf.\
  figure~\ref{fig:sketch}~c).\footnote{An event with exactly three
    jets can have $\Dphi = 2\pi/3$ only in a ``Mercedes Star''
    configuration, where the jets have $p_\mathrm{T1} = p_\mathrm{T2}
    = p_\mathrm{T3}$ and $\Delta\phi_{1,2} = \Delta\phi_{1,3} =
    \Delta\phi_{2,3} = 2\pi/3$.  If the two jets leading in $\pt$ in a
    three-jet event (with $p_\mathrm{T1} \ge p_\mathrm{T2} \ge
    p_\mathrm{T3}$) had $\Dphi < 2\pi/3$, the vector sum of their
    transverse momenta could only be balanced, if the third jet had
    $p_\mathrm{T3} > p_\mathrm{T2}$, which would, however, contradict
    the assumption that $p_\mathrm{T2} \ge p_\mathrm{T3}$.}
  Therefore, for $\yinit = \infty$, the dijet cross section for $\Dphi
  < 2\pi/3$ is a four-jet quantity, meaning that the lowest order pQCD
  contributions are from the four-jet tree-level matrix elements.

\item {\bf Kinematic constraints for a limited \boldmath $y$ region,
    $\yinit < \infty$} \\
  If the $y$ region for the initial jet selection is limited, it is
  possible that the two jets, selected for the dijet system, are not
  the two jets leading in $\pt$ of the whole event. Table~\ref{tab:evt} gives
  an example for the ATLAS scenario, in which the leading jet in
  the event has $|y| > \yinit$. In this case, the dijet system is made
  of the second and third leading jets, which are the two
  highest $\pt$ jets inside the limited $y$ region. Since there is no
  kinematic constraint for the azimuthal angular separation between the
  second and third leading jet, the
  region $\Dphi < 2\pi/3$ is also populated by three-jet final states.
  If such configurations are not prohibited by other phase space
  constraints, the dijet cross section for $\Dphi < 2\pi/3$ is a
  three-jet quantity.

\end{itemize}

\begin{table}[tbp]
  \centering
  \begin{tabular}{|c|c|c|c|}
    \hline
    & jet 1  & jet 2 & jet 3 \\
    \hline
    $\pt$ (GeV) & $405$ & $401$ & $101$ \\
    $y$  & $2.805$ & $-0.75$ & $-0.75$ \\
    $\phi$ (radians) &  $0.000\cdot\pi$&  $0.920\cdot\pi$&  $1.448 \cdot\pi$ \\
    \hline
  \end{tabular}
  \hskip9mm
  \begin{tabular}{|c|c|}
    \hline
    $\Delta\phi_{2,3}$ (radians) &   $0.528\cdot\pi$  \\
    $M_{\mbox{\scriptsize 3-jet}}$ (TeV) & 2.745 \\
    \hline
    $x_1$ (for $\sqrt{s}=7\,$TeV) & 0.990 \\
    $x_2$ (for $\sqrt{s}=7\,$TeV) & 0.155 \\
    \hline
  \end{tabular}
  \caption{\label{tab:evt}The topology of an exclusive three-jet event,
    with the jet variables $\pt$, $y$, and $\phi$ (left)
    and the event quantities $\Delta\phi_{2,3}$,
    three-jet invariant mass $M_{\mbox{\scriptsize 3-jet}}$,
    and the momentum fractions $x_1$ and $x_2$
    for a center-of-mass energy of $\sqrt{s}=7\,$TeV.
    In this event, the highest $\pt$ jet is produced at large rapidity.
    If the dijet selection is restricted to jets with $|y| < \yinit =2.8$
    (as in the ATLAS scenario, see text),
    the selected dijet system does not include the highest $\pt$ jet.
    This enables the azimuthal angular separation of the jets in the dijet system
    (here, $\Dphi$ is determined by the azimuthal angle between the second and
    the third jet, $\Delta\phi_{2,3}$)
    to fall below the limit of $\Dphi = 2\pi/3$.}
\end{table}

\begin{figure}
  \centering
  \includegraphics[scale=1.1]{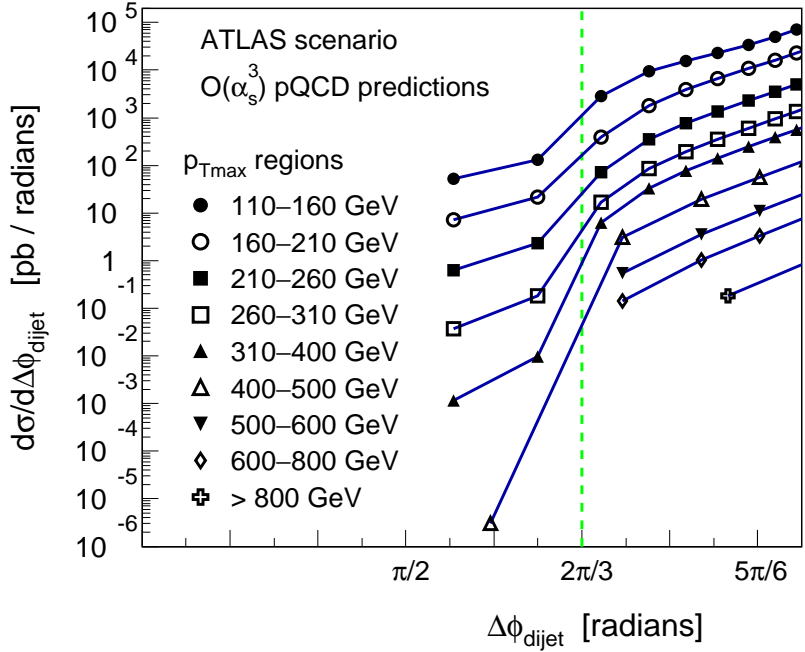}
  \caption{\label{fig:atllo}The pQCD predictions of order
    $\ord(\asss)$ for the dijet differential cross section ${\rm
      d}\sigma / {\rm d}\Dphi$, as a function of $\Dphi$ in different
    regions of $\ptmax$ for all analysis bins of the ATLAS measurement.
    The figure demonstrates that the $\ord(\asss)$ contributions
    to bins with $\Dphi < 2\pi/3$ and $\ptmax < 500\,$GeV are small
    but non-zero.}
\end{figure}

It depends on the requirements on $\yinit$, $(\ptmax/\sqrt{s})$, and
$(\ptmin/\sqrt{s})$, whether a leading jet is kinematically allowed
outside the region $|y|<\yinit$ and, as a consequence, three-jet
configurations can populate the region of $\Dphi < 2\pi/3$.
This can be tested using a cross section calculation based on
tree-level $2 \rightarrow 3$ matrix elements as \eg in
\nlojet~\cite{Nagy:2003tz,Nagy:2001fj}. We have used \nlojet to
compute the dijet differential cross section ${\rm d} \sigma_{\rm
  dijet}/{\rm d} \Dphi$ for all three scenarios.
The results for the ATLAS scenario are shown in figure~\ref{fig:atllo}
and it is observed that up to and including the $\ptmax$ region of
400--500\,GeV, the dijet differential cross section ${\rm d}
\sigma_{\rm dijet}/{\rm d} \Dphi$ receives non-zero contributions at
$\Dphi < 2\pi/3$ from three-jet final states. Therefore, in the ATLAS
scenario, ${\rm d} \sigma_{\rm dijet}/{\rm d} \Dphi$ is a three-jet
quantity for all $\Dphi$ in the $\ptmax$ regions with $\ptmax <
500\,$GeV.
Only in the higher $\ptmax$ regions it becomes a four-jet quantity. In
those regions, however, ATLAS has not published any measurement for
$\Dphi < 2\pi/3$.

Like ATLAS, the CMS scenario also has a limited $y$ region for the
initial jet selection, with lower requirements for $\ptmax$ and
$\ptmin$, but with a larger value of $\yinit = 5.0$.  We have
computed ${\rm d} \sigma_{\rm dijet}/{\rm d} \Dphi$ for the CMS
scenario as well and find that in all $\ptmax$ regions the $2 \rightarrow 3$
tree-level predictions for ${\rm d} \sigma_{\rm dijet}/{\rm d} \Dphi$
are zero for $\Dphi < 2\pi/3$.  In other words, in both the CMS and
the \DO scenarios ${\rm d} \sigma_{\rm dijet}/{\rm d} \Dphi$ is a
four-jet quantity for $\Dphi < 2\pi/3$.

\noindent We summarize our findings as follows:
\begin{itemize}
\item%
  The denominator of $P$, $\sigma_{\rm dijet}$, is the inclusive dijet
  cross section, which is a two-jet quantity in all scenarios.

\item%
  For $\Dphi \ge 2\pi/3$, the numerator of $P$, ${\rm d}\sigma_{\rm
    dijet}/{\rm d}\Dphi$, is a three-jet quantity in all scenarios.

\item%
  For $\Dphi < 2\pi/3$, the numerator of $P$ is a four-jet quantity, if
  the initial $y$ region is unlimited ($\yinit = \infty$) as in the
  \DO scenario, or if the $\yinit$ and $\pt$ requirements prohibit
  the two jets with the highest $\pt$'s in an event from having $|y| > \yinit$,
  as in the CMS scenario.

\item%
  If the $\yinit$ and $\pt$ requirements allow one of the two jets
  leading in $\pt$ to have $|y| > \yinit$, then the numerator of $P$ is a
  three-jet quantity for all $\Dphi$. This is the case in the ATLAS
  scenario for the $\ptmax$ regions up to 400--500\,GeV in $\ptmax$.

\end{itemize}

\section{Perturbative QCD calculations for cross section ratios
\label{sec:pqcd}}

The pQCD prediction for a ratio $R$ of two cross sections $\sigma_A$
and $\sigma_B$ in a given relative order of $\as$ (\eg LO or NLO) can
be computed from the ratio of the pQCD predictions for $\sigma_A$ and
$\sigma_B$. For this purpose, both must be computed at the
same {\em relative}\/ order,
which is not necessarily the same absolute order in $\as$.
A LO result is then given by $R_{\rm LO} = \sigma_A^{\rm LO} /
\sigma_B^{\rm LO}$ and a NLO result by $R_{\rm NLO} = \sigma_A^{\rm
  NLO} / \sigma_B^{\rm NLO}$. If numerator and denominator are
calculated in different relative orders, cancellation effects between
theoretical uncertainties can be compromised leading to
an artificially increased renormalization scale dependence
of the results as discussed with respect to jet shapes in
sections~3.1 and~4 of reference~\cite{Seymour:1997kj}.

For two-jet quantities, the LO and NLO pQCD predictions are given by
calculations to order $\ord(\ass)$ and $\ord(\asss)$, respectively.
For each additional jet required for the final state, the respective
powers of $\as$ increase by one, so that for example
the LO (NLO) predictions for three-jet quantities are given by pQCD
calculations to order $\ord(\asss)$ ($\ord(\assss)$).
Combined with the findings from section~\ref{sec:ps}, we obtain the
rules for the calculation of the LO and NLO results for the quantity
$P$ in the three scenarios and in the different regions of $\Dphi$.
These rules are listed in table~\ref{tab:calc} and compared to
the computational procedures applied in the experimental
publications~\cite{Abazov:2004hm,Khachatryan:2011zj,daCosta:2011ni}.
The theory results published by \DO and CMS for $\Dphi < 2\pi/3$
and labeled ``NLO'' in
references~\cite{Abazov:2004hm,Khachatryan:2011zj} are inconsistent,
because they mix relative orders for the numerator (LO) and
denominator (NLO).
Replacing the NLO result for the denominator (in $\ord({\asss})$) by
the corresponding LO ($\ord({\ass})$) provides the correct LO result
for $P$ below $\Dphi = 2\pi/3$.
Alternatively, the correct NLO results at $\Dphi < 2\pi/3$ can be
obtained by replacing the four-jet LO ($\ord({\as^4})$) results by
results based on the four-jet matrix elements at NLO pQCD
($\ord({\as^5})$), which have become available in the last
years~\cite{Bern:2011ep,Badger:2012pf}.

\begin{table}[tbp]
\renewcommand{\arraystretch}{1.1}
\small
\centering
\begin{tabular}{|c|c|c|c|c|c|}
\hline
scenario & $\Dphi$ range & order for $P$  &  numerator & denominator  &
            used in publication \\
\hline
\multirow{2}{*}{ATLAS}          &   \multirow{2}{*}{all $\Dphi$}
                       & LO $\ord(\as)$ & $\ord(\asss)$  & $\ord(\ass)$ &
             the published LO and \\
\cline{3-5}
              &             & NLO $\ord(\ass)$ & $\ord(\assss)$ &  $\ord(\asss)$  &
                NLO are correct \\
\hline
  & \multirow{2}{*}{$ \ge 2\pi/3$}
                       & LO $\ord(\as)$ & $\ord(\asss)$ & $\ord(\ass)$ &
                    the published LO and \\
\cline{3-5}
\DO        &             & NLO $\ord(\ass)$ & $\ord(\assss)$ & $\ord(\asss)$ &
               NLO are correct  \\
\cline{2-6}
and   &  \multirow{4}{*}{$ < 2\pi/3$}
                       & \multirow{2}{*}{LO $\ord(\ass)$} &
                   \multirow{2}{*}{$\ord(\as^4)$} &  \multirow{2}{*}{$\ord(\ass)$} &
            numerator: LO $\ord(\as^4)$  \\
 CMS                 & &  & & &
                      denominator: NLO $\ord(\asss)$   \\
\cline{3-5}
              &          & \multirow{2}{*}{NLO $\ord(\asss)$} &
     \multirow{2}{*}{$\ord(\as^5)$}  &  \multirow{2}{*}{$\ord(\asss)$} &
    (inconsistent, using \\
             & &  & &  &
                             mixed relative orders)\\
\hline
\end{tabular}
\caption{\label{tab:calc}Correspondence between absolute orders in $\as$
  in the calculations of numerator and denominator
  and the relative order in the quantity $P$.
  The right column comments on the calculations used
  in the experimental publications.
}
\end{table}

\begin{figure}
  \centering
  \includegraphics[scale=1]{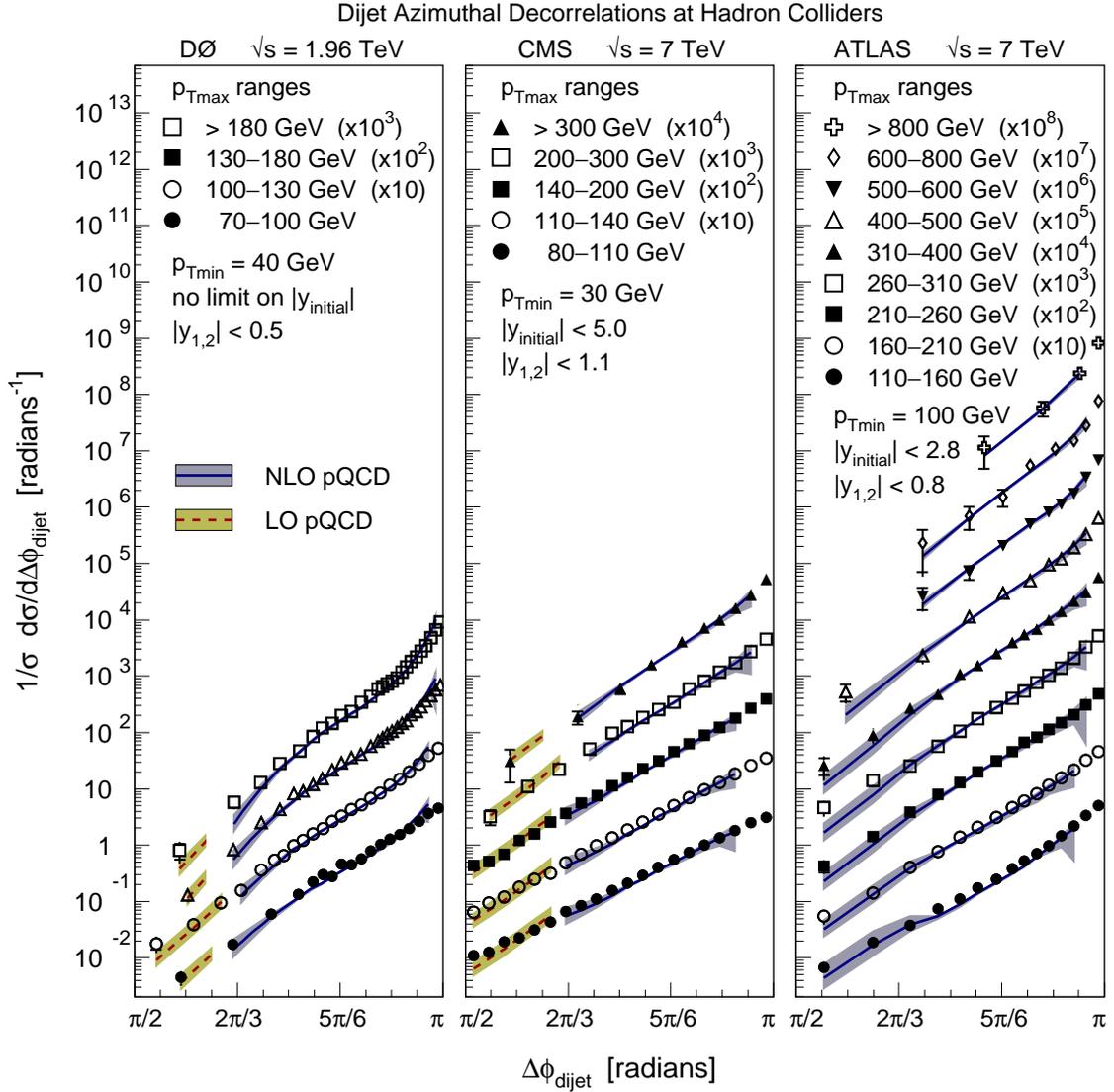}
  \caption{\label{fig:data}Measurements of
    dijet azimuthal decorrelations at hadron colliders from
    the \DO, CMS, and ATLAS experiments (from left to right) are
    displayed as a function of the azimuthal opening angle $\Dphi$ of
    the dijet system for different requirements of the leading jet
    $\pt$ (different markers). The measurements are compared
    to theoretical predictions based on NLO (solid lines) or LO pQCD
    (dashed lines), depending on whether the measured quantity is a
    three-jet or four-jet variable, respectively. The scale
    dependence of the pQCD calculation is indicated by the shaded
    areas.}
\end{figure}

\section{Results\label{sec:results}}

Following the prescriptions in table~\ref{tab:calc}
we have computed the LO and NLO pQCD predictions for $P$
in the \DO, CMS, and ATLAS scenarios in the different $\Dphi$ regions.
For comparison, we also derive the inconsistent ``mixed-order''
results for $P$ as published by \DO and CMS.

All calculations are made in the
$\overline{\rm MS}$-scheme~\cite{Bardeen:1978yd} and for five massless quark
flavors, using \nlojet~\cite{Nagy:2003tz,Nagy:2001fj} interfaced to
\fastnlo~\cite{Kluge:2006xs,Britzger:2012bs}.
The results are obtained for renormalization and factorization scales
of $\mu_R = \mu_F = \ptmax$, with the MSTW2008NLO~\cite{Martin:2009iq}
parameterization of the parton distribution functions of the proton,
and with $\as$ evolved from a value of $\asmz = 0.120$
according to the two-loop solution of the renormalization
group equation.
The uncertainty due to the scale dependence is computed from the
variations of the ratio $P$ for correlated variations of the scales in
the numerator and denominator of $\mu_R = \mu_F = \ptmax/2$ and $\mu_R
= \mu_F = 2\,\ptmax$.
The ATLAS collaboration has published
non-perturbative corrections~\cite{Whalley:1989mt,Buckley:2010jn},
which are applied to the pQCD results to get the final theory prediction.
These corrections are typically below 1\% and never larger than 3\%.
The \DO and CMS collaborations have not provided non-perturbative corrections.
In these cases, the pQCD results are directly compared to the
data.\footnote{In reference~\cite{Wobisch:2004ru} non-perturbative
corrections for the \DO results are shown to be typically below 2\% and
never larger than 4\%.
In the CMS publication~\cite{Khachatryan:2011zj}
the non-perturbative corrections are quoted to
vary between $-13\%$ at $\Dphi = \pi/2$ and $+4\%$ at $\Dphi = \pi$.}
The theoretical calculations in this study differ slightly from the
calculations used in the CMS and \DO publications due to different choices
of the parton distribution functions and $\asmz$.
Furthermore, the \DO collaboration chose different
renormalization and factorization scales of $\mur=\muf=\ptmax/2$,
and the CMS collaboration applied non-perturbative corrections.
For the purpose of the following discussion,
these differences are negligible.

\begin{figure}
  \centering
  \includegraphics[scale=1]{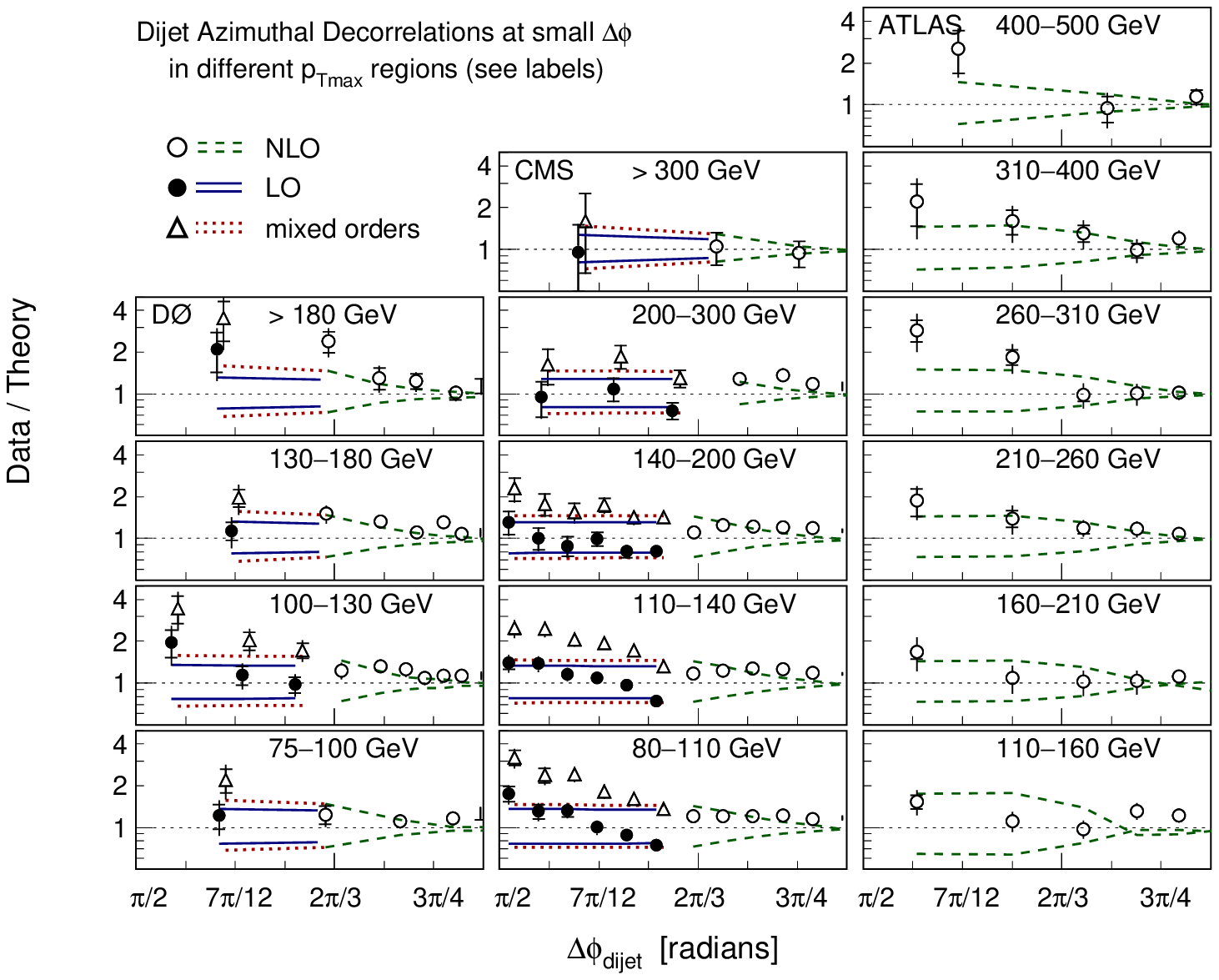}
  \caption{\label{fig:ratio}Ratios of data from different experiments
    (columns) to fixed-order predictions as a function of $\Dphi$,
    from low $\ptmax$ (bottom) to high $\ptmax$ (top). The ratios are
    shown in different regions of $\Dphi$ for the pQCD predictions at
    NLO (open circles) and LO (full circles), and also for the case
    of mixing different orders in numerator and denominator
    (triangles) for $\Dphi < 2\pi/3$. For better visibility the full
    circles have been slightly shifted towards smaller values of $\Dphi$.
    The scale dependence of the different pQCD calculations is
    indicated by the corresponding lines.}
\end{figure}

The experimental results from the \DO, CMS, and ATLAS
measurements are displayed in figure~\ref{fig:data} over the entire
$\Dphi$ range. The data are compared to theory at NLO or LO,
depending on the $\Dphi$ range and the scenario.
Over the whole range of $\ptmax$ and $\Dphi$, the
theoretical predictions are in agreement with the data, except for the
ATLAS data at small $\Dphi$.

The region of small $\Dphi$, including the transition at $\Dphi =
2\pi/3$ and the effects of the inconsistent mixed-order predictions,
are further investigated in the following.
Figure~\ref{fig:ratio} shows the ratios of data over the different
theory predictions for $\Dphi \lesssim 3\pi/4$. The ratios are
computed for the NLO results, the LO results, and the inconsistent
results from mixed relative orders.  Also shown are the uncertainty
bands due to the scale dependence of the different theoretical
calculations.
For $\Dphi > 2\pi/3$, in all scenarios the NLO pQCD predictions are
compared to the data.  For $\Dphi > 3\pi/4$, these give a good
description of the data within scale uncertainties, which are below
5--10\%.  In the range $ 2\pi/3 < \Dphi < 3\pi/4$, the $\ord{(\as^4)}$
(\ie three-jet NLO) calculation for the numerator is running out of
phase space for three-jet final states as $\Dphi \rightarrow 2\pi/3$.
This causes the $\ord(\assss)$ calculation to effectively become a
four-jet LO calculation.  In this $\Dphi$ range the NLO prediction
still describes the data, but with an increasing scale dependence
of up to 30\% as $\Dphi \rightarrow 2\pi/3$.

For the CMS and \DO scenarios at $\Dphi < 2\pi/3$, we first focus on
the inconsistent mixed-order calculations as published by the
experiments.
Figure~\ref{fig:ratio} shows that over most of the range (and in
particular towards lower $\Dphi$) these predictions are significantly
below the data even outside their large scale dependence, and they do not
describe the $\Dphi$ dependence of the data.
Compared to the inconsistent mixed-order calculations, the
correct LO predictions have a significantly reduced scale dependence,
and they give a much better description of the data. While they still
do not reproduce the $\Dphi$ dependence, almost all individual data
points agree with the LO prediction within the reduced scale uncertainty.

Although, for the ATLAS scenario the pQCD predictions for $\Dphi <
2\pi/3$ are technically still of NLO, their scale dependence is as
large as that of the mixed-order predictions for the CMS scenario, and
the description of the data by both are equally poor.

\section{Recommendations for future measurements\label{sec:future}}

In section~\ref{sec:pqcd} we pointed out that for the numerator
of $P$ in the ATLAS scenario the three-jet NLO cross section calculations
formally are of NLO also for $\Dphi < 2\pi/3$.
The results presented in section~\ref{sec:results}, however, demonstrate
that these NLO predictions exhibit a larger scale dependence and that they
give a worse description of the data than the LO predictions
for the \DO and CMS results. The difference between the ATLAS
and the \DO and CMS scenarios was traced back to
the choice of $\yinit$ in the dijet selection as explained in
section~\ref{sec:ps}.
In contrast to the \DO and CMS scenarios, the kinematic constraints
in the ATLAS scenario do allow $2 \rightarrow 3$ processes to give small,
but non-zero contributions to the dijet cross
section for $\Dphi < 2\pi /3$.
Therefore, in this $\Dphi$ range, while formally being
a NLO pQCD prediction, the $\ord(\assss)$ calculation
for the numerator effectively is only a LO prediction,
since the $\ord(\asss)$ terms contribute less than one percent.
This ``formally NLO but effectively LO'' calculation for the
numerator exhibits the typical large scale dependence of a LO
calculation while the NLO predictions for the denominator have a
reduced scale dependence, as typical for NLO calculations. As a consequence,
the NLO prediction for the ratio $P$ has a scale
dependence, which is similar to that of the mixed-order calculations
and larger than that of the LO predictions for the \DO and CMS scenarios.

Therefore, we strongly recommend that future measurements of dijet
azimuthal decorrelations use values of
$\yinit$ that, together with the $\ptmin$ and $\ptmax$ requirements,
do not leave any phase space for $2 \rightarrow 3$ processes below
$\Dphi = 2\pi/3$.
Technically, this can be investigated by using a
phase space generator or a three-jet pQCD LO calculation for the
numerator of $P$.

\section{Summary and conclusion}

Measurements of dijet azimuthal decorrelations at hadron colliders
continue to be a testing ground for pQCD predictions at higher orders,
beyond what is probed in inclusive jet and inclusive dijet production.
In particular in the phase space region of $\Dphi < 2\pi/3$, dijet
azimuthal correlations are sensitive to the dynamics of final states
with four or more jets. In all previous publications of azimuthal
decorrelations, based on the quantity $P= (1/\sigma_{\rm dijet}) \cdot
({\rm d} \sigma_{\rm dijet}/{\rm d} \Dphi)$, this region was
poorly described by theoretical predictions. In this article we
have identified two reasons for this shortcoming. 

In the publications by \DO~\cite{Abazov:2004hm} and
CMS~\cite{Khachatryan:2011zj}, the poor theoretical description of the
data is related to the inconsistent mixing of different relative
orders in $\as$ in the predictions for the ratio $P$.
We have performed a consistent LO calculation by computing both, numerator
and denominator, at LO\@.
This correct LO pQCD prediction not only exhibits a smaller scale dependence,
but also gives a better description of the experimental data for
$\Dphi < 2\pi/3$.

The improvement due to the consistent LO calculation can, however,
only be achieved for definitions of the dijet phase space that ensure
the two jets of the dijet system to be also the two leading $\pt$ jets
in the events.
We strongly recommend for future measurements of
dijet azimuthal decorrelations at small $\Dphi$
to perform the initial dijet selection accordingly.

If this is taken into account,
the future usage of four-jet NLO calculations will
provide NLO pQCD predictions for the whole $\Dphi$ range, extending
precision phenomenology for dijet azimuthal decorrelations to the
region $\Dphi < 2\pi/3$.
Since in this $\Dphi$ region the quantity $P$ is proportional to $\as^2$,
future measurements with higher statistical precision can also be used
for novel $\as$ determinations.
This recommendation also applies to measurements of dijet azimuthal
decorrelations based on the quantity
$\Rdphi$~\cite{Wobisch:2012au,Abazov:2012jhu} when this is measured
for $\Dphimax \le 2\pi/3$.

\acknowledgments

We thank our colleagues in the ATLAS, CMS, and D0 collaborations
for many fruitful discussions.
The work of M.W. is supported by grants DE-FG02-10ER46723 and
DE-SC0009859 from the U.S. Department of Energy.
M.W. also wishes to thank the Louisiana Board of Regents Support Fund
for the support through the Eva J. Cunningham Endowed Professorship.

\bibliography{ondphi}

\end{document}